\documentclass[useAMS,usenatbib]{mn2e}

\title[Modified Chaplygin Gas ...  ]{Modified Chaplygin Gas and Constraints on its B parameter from CDM and UDME Cosmological models
}
\author[P. Thakur, S. Ghose and B. C. Paul]
  {P. Thakur,$^1$ \thanks{ Electronic mail : prasenjit \textunderscore thakur1 @yahoo.co.in}
  S. Ghose,$^2$ \thanks{ Electronic mail : souviknbu@rediffmail.com} 
  B. C. Paul,$^2$ \thanks{ Electronic mail : bcpaul@iucaa.ernet.in }
   \\
  $^1$Physics Department, Alipurduar College \\ 
      Dist. : Jalpaiguri, Pin : 736122, West Bengal, India
\\
  $^2$Physics Department, North Bengal University \\
      Dist. : Darjeeling, Pin : 734 013, West Bengal, India \\}

\date{}
\begin{document}

%\date{Accepted 1988 December 15. Received 1988 December 14; in original form 1988 %October 11}

\pagerange{\pageref{firstpage}--\pageref{lastpage}} \pubyear{2008}

\maketitle

\label{firstpage}

\begin{abstract}

We study  Modified Chaplygin Gas (MCG) as a candidate for dark energy and predict the values of parameters of the gas for a physically viable cosmological model. The equation of state of MCG ($p=B \rho - \frac {A}{\rho^\alpha} $) involves three parameters: $B$, $A$ and $\alpha$. The permitted values of these parameters are determined with the help of dimensionless age parameter ($H_{o}t_{o}$) and $H(z)-z$ Data. Specifically we study the allowed ranges of values of B parameter in terms of  $\alpha$ and $A_{s}$ ($A_{s}$ is defined in terms of the constants in the theory). We explore the constraints of the parameters in Cold Dark Matter(CDM) model and UDME(Unified Dark Matter Energy) model respectively. 

\end{abstract}


\section{Introduction}

Recent cosmological observations, such as high redshift surveys of SNe Ia (Perlmutter {\it et al.} 1997a; Permutter {\it et al.}  1997b; Riess {\it et al.}  1998; Tonry {\it et al.}  2003), CMBR (Melchiorri {\it et al.}  2000; Lange 2001; Jaffe  {\it et al.}  2001; Netterfield {\it et al.}  2002; Halverson {\it et al.}  2002), WMAP (Briddle {\it et al.}  2003; Bennet {\it et al.}  2003; Hinshaw  {\it et al.}  2003; Kogut {\it et al.}  2003; Spergel {\it et al.}  2003) predict that our present universe is  passing through an accelerated phase of expansion preceeded by a period of deceleration. It is known that the ordinary matter and fields of the standard model are not sufficient to accommodate the present phase of acceleration (preceede by deceleration). Consequently a modification of matter sector of the Einstein Gravity is essential to incorporate the recent prediction from observational cosmology. The notion of new type of matter has come up which must have negative pressure. Recent astronomical data when interpreted in the context of Big Bang Model have provided some interesting information about the composition of the universe. The analysis reveals that our universe is spatially flat and consists of 70 percent dark energy with negative pressure, remaining 30 percent dust matter (cold dark matter plus baryons), and negligible radiation. It has been predicted that the dark energy may be responsible for the present acceleration of our universe.

The most simple candidate for these uniformly distributed (i.e unclustered) dark energy is considered to be in the form of vacuum energy density or cosmological
constant ($\Lambda$). The model with cosmological constant is entangled with (i) fine tuning problem (present amount of the dark energy is so small compared with fundamental scale) and (ii) coincidence problem (dark energy density is comparable with critical density today). Alternatively the other choices are (i) a light homogeneous scalar field  $\phi$, whose effective potential $V(\phi)$ leads to an
accelerated phase at a later stage of the universe (Caldwell {\it et al.} 1998; Sahani {\it et al.} 2000), (ii) a X-matter component, which is characterized by an equation of
state $p=\omega\rho$, where, $-1\leq\omega<0 $ (Peebles and Ratra 2002), (iii) effects from extra dimensions (Sahni 2002; Lui 2002), (iv) an exotic fluid, Chaplygin Gas etc..

The motivation of the paper  is to obtain cosmological model constraining Chaplygin Gas taking into account observational facts. Chaplygin Gas was first introduced in aerodynamics in 1904. Recently it has been shown that Chaplygin Gas may be useful for describing  dark energy because of its negative pressure. Although it has positive energy density it carries a negative pressure for which it is referred to as   exotic fluid. In the context of string theory the Chaplygin gas emerges from the dynamics of a generalized d-brane in a (d+1,1) spacetime. It can be described by a complex scalar field which is obtained from a generalized Born-Infeld action. The equation of state is given by
\begin {equation}
p=-\frac{A}{\rho}
\end{equation}
where $A$ is a positive constant, $p$ and $\rho$ are pressure and density respectively. Subsequently a modified form of the equation of state (Bento {\it et al.} 2002; Bilic, Tupper 2001) of the form
\begin{equation}
p=-\frac{A}{\rho^\alpha}
\end{equation}
with $ 0 < \alpha \leq 1 $ which is known as Generalized Chaplygin Gas (GCG). It has two free parameter $A$ (positive), $\alpha$. In GCG model, at low energy density, the fluid
pressure is negative and constant while at high energy density it behaves almost like pressure less fluid. Thus it smoothly interpolates between a non-relativistic matter phase in the past and a negative pressure dark energy regime at late times.

Recently a modified form of GCG has been considered in cosmology (Liu and Li 2005). The modified Chaplygin gas (MCG) is more general and contains three free parameters. The idea is  to interpolate states of standard fluids at high pressures and at high energy densities to a constant negative pressure at low energy densities (Debnath {\it et al.} 2004). In addition it covers whole aspects of GCG. This model accommodates  consistent (i) Gravitational lensing test (Silva {\it et al.}  2003; Dev {\it et al.}  2004), (ii) Gamma-ray bursts (Bertolami and Silva 2006).
The equation of state for this Modified Chaplygin Gas is given by
\begin{equation}
p=B\rho-\frac{A}{\rho^\alpha}
\end{equation}
where $A$, $B$, $\alpha$ are arbitrary constants
with $0 \leq \alpha \leq 1 $. As there are three free parameters, unlike GCG, we look for a suitable range of B parameter for MCG for a viable cosmological model accommodating the observational evidences. The parameters are determined by 
(i) Considering a  dimensionless age parameter $H_{o}t_{o}$ (Dev {\it et al.} 2002) and
(ii) $H(z)-z$ Data analysis (Wu and Yu 2006).

We perform the following analysis:

Case 1. The age parameter ($H_{o}t_{o}$ ) is dimensionless and a constant irrespective of the model we are considering. For simplicity we chose its standard value to be 0.95 (ignoring error). Imposing this constant age parameter we determine the effective ranges of values of the free parameters in this model. As the parameters have some preferred range of values, one can ultimately constrain one of them, in particular the matter part B.

Case 2. Using ($H(z)-z$) Data we further verify the validity of the constraints on the parameters obtained in case 1. We use Hubble parameter vs redshift relation given in Table 1. The $\chi^2$ minimization technique has been used in this process.
There are 9 data points of $H(z)$ at redshift $z$ used to constrain the MCG model.

  \begin{table}
  \begin{minipage}{140mm}
  \caption{}
  \begin{tabular}{l|c|r}
  \hline
  {\it z Data} & H(z) & $\sigma$ \\
  \hline
    0.09 & 69  & $ \pm $ 12.0	 \\
   0.17 & 83  & $ \pm \;\; $ 8.3  \\
   0.27 & 70  & $ \pm $ 14.0   \\
   0.40 & 87  & $ \pm $ 17.4 \\
   0.88 & 117 & $ \pm $ 23.4 \\
   1.30 & 168 & $ \pm $ 13.4 \\
   1.43 & 177 & $ \pm $ 14.2 \\
   1.53 & 140 & $ \pm $ 14.0   \\
   1.75 & 202 & $ \pm $ 40.4 \\

\hline
\end{tabular}
\end{minipage}
\end{table} 

We investigate both CDM and UDME (Unified Dark Matter Energy) models in next sections.UDME model refers to the model in which the Modified Chaplygin gas(MCG) represent dark matter and dark energy as a whole, where the total energy density comprises of radiation, baryon and MCG energy density. In case of Cold dark matter (CDM) model the constituents of our universe are radiation, CDM and MCG.

This paper is organized as follow : In section II we present the relevant field equations and introduce Hubble parameter and deceleration parameter respectively. In section III and IV we explore values of $B$ parameters using the age parameter and the observed $H(z)-z$ data respectively. In section V we checked viability of the models from union compilation data. In section VI we summerise the result. 

\section[]{FIELD EQUATIONS, DECELERATION PARAMETER \\}

Let us now consider the Friedmann-Robertson-Walker (FRW)line element (c=1)

\begin{equation}
ds^{2} = - dt^{2} + a^{2}(t) \left[ \frac{dr^{2}}{1- k r^2} + r^2 ( d\theta^{2} + sin^{2} \theta \;
d  \phi^{2} ) \right]
\end{equation}
where  $k=0,\pm1$ is the curvature parameter in the spatial section $a(t)$ is the scale factor of the universe, $r,\theta,\phi$ are the comoving co-ordinates.
The energy conservation equation is given by 
\begin{equation}
\frac{d\rho}{dt} + 3 H (\rho + p) = 0 ,
\end{equation}
where $p$, $\rho$, $H$ are  pressure, energy density, Hubble parameter respectively.
Using eq. (3) in eq. (5) we obtain the expression for the energy density
of MCG with the scale factor of the universe, which is given by,
\begin{equation}
\rho = \left[\frac{A}{1+B} + \frac{C}{a^{3n}} \right]^{\frac{1}{1+
\alpha}}
\end{equation}
where  $C$ is an arbitrary  constant and we denote $  (1 + B) ( 1 +
\alpha) = n $.
Equation (6) can be rewritten as 
\begin{equation}
\rho=\rho_o\left[A_S+\frac{1-A_S}{a^{3n}}\right]^{\frac{1}{1+
\alpha}}
\end{equation}
where $z$ is redshift parameter, $ A_S = \frac{A}{1+B}\frac{1}{\rho_o^{\alpha +1}}$, $\frac{a}{a_0}=\frac{1}{1+z}$ and we chose $a_0=1$ for convenience. It reduces to GCG model when we set $B=0$ in the above equation.
The Friedmann`s equation becomes

\begin{eqnarray}
H(z)=H_0[\Omega_{r0}(1+z)^4+\Omega_{j0}(1+z)^3
+  \nonumber \\
\;\;\;\;\;(1-\Omega_{r0}-\Omega_{j0})
[(A_s+(1-A_s)(1+z)^{3n})^{\frac{1}{1+\alpha}}]]^{\frac{1}{2}}.
\end{eqnarray}
The above equation can be rewritten in terms of $a$ as
\begin{eqnarray}
H(a)=H_0 [\frac{\Omega_{r0}}{a^4}+\frac{\Omega_{j0}}{a^3}\nonumber \\
\;\;\;\;\;+(1-\Omega_{r0}-\Omega_{j0})( A_s+\frac{1-A_s}{a^{3n}})^{\frac{1}{1+\alpha}} ] ^{\frac{1}{2}}
\end{eqnarray}
where $j = m$ for CDM model and $j = b$ for UDME model. The above reduces to GCG model when one sets $B=0$. The deceleration parameter ($q_0=-(\frac{a\ddot{a}}{\dot{a}^{2}})_{t_0}$) at the present time can be written as
\begin{equation}
q_0=\frac{3}{2}\left[\frac{\Omega_{j0}+\frac{4}{3}\Omega_{r0}+
(1+B)(1-\Omega_{j0}-\Omega_{r0})(1-A_S)}{\Omega_{j0}+\Omega_{Cg_{0}}+\Omega_{r0}}\right]-1
\end{equation}
where $j=m$ for CDM model and $j=b$ for UDME model. The deceleration parameter can be estimated both in CDM and UDME model.
For a flat universe we have $ \Omega_{j0}+\Omega_{Cg_{0}}+\Omega_{r0}=1$ which will be used to measure the parameters in the next section. In the above $\Omega_{Cg_{0}}$ represents the present day Modified Chaplygin gas energy density, $ \Omega_{j0}$ is the present energy density of either Cold dark matter (in CDM model) or baryonic energy density (in UDME model) and $\Omega_{r0}$ represents the present radiation energy density of our universe.

\section{AGE OF OUR UNIVERSE  AS A CONSTRAINING TOOL}

Using the definition of the age parameter
\begin{equation}
t_{0}=\int_{0}^{1}  \left[\frac{da}{aH(a)} \right ]
\end{equation}
where $\frac{a}{a_0}=\frac{1}{1+z}$ and $H(a)$ is given by eqn. (9). The predicted age of the universe in  MCG model becomes

\begin{equation}
t_{0}=\frac{1}{H_{0}}\int_{0}^{1} \left[\frac{da}{a f \left(a,\Omega_{j0},\Omega_{r0},A_S,B,\alpha \right )}\right]
\end{equation}
with
\begin{equation}
f(a,\Omega_{j0},\Omega_{r0},A_S,B,\alpha)=\frac{H(a)}{H_{0}}.
\end{equation}
We consider here from experimental facts the value  $H_{o}t_{o}=0.95\;$. Although it has some error 
limit in both sides, we take this value as standard. From the constancy of this parameter, we derive constraints on the parameters of the theory. For a given value of alpha we plot variation of $A_{s}$ with $B$. We note the following :

\input{epsf}
\begin{figure}
\begin{center}
\epsffile{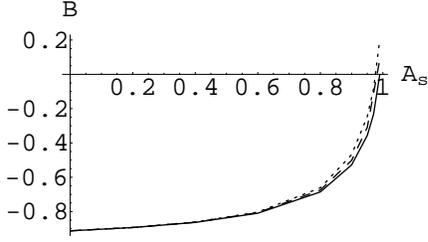}
\caption {CDM,Dotted line for $\alpha = 0.0$,Dashed line for $\alpha = 0.2$,Thin line for $\alpha=0.39$ }
\end{center}
\end{figure}

Fig.1 : shows variation of $B$ with $A_{s}$ for $\alpha=0,0.20,0.39$ by dotted, dashed and thin lines respectively in CDM model.
It is evident that as the value of $A_{S}$ approaches 1 (0.97 to 1) for $0\;\leq\; \alpha\; \leq\;0.39\;$ the B parameter picks up positive values with a maximum 0.20.

\input{epsf}
\begin{figure}
\begin{center}
\epsffile{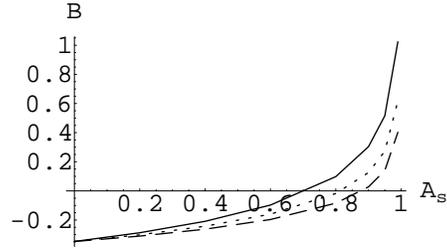}
\caption { UDME,Thin line for $\alpha = 0.0$,Dotted line for $\alpha = 0.5$,Dashed line for $\alpha = 1.0$ }
\end{center}
\end{figure}

Fig. 2: shows variation of $B$ with $A_{s}$ for different $\alpha$  $\alpha$=0,0.5 and 1 with thin, dotted and dashed lines respectively in UDME. In this case as the value of $A_{S}$ is increased from 0.7 to 1 it is evident that the B parameter picks up
positive value up to a maximum 1.02 for $0\;\leq\; \alpha\; \leq\;1\;$ in UDME model.

The set of curves shown in fig.1 and fig.2 is useful to determine the range of values of $B$ for both CDM and UDME models respectively. We note that in CDM model $B$ lies between 0 to 0.20, where as in UDME model  $B$ lies between 0 to 1.02. Moreover, in CDM model $B$ is positive when $0 \leq \alpha \leq 0.39$ and $A_s$ between 0.97 to 1. In UDME we note that $B$ is positive for $0 \leq \alpha \leq 1$ and $A_s$ between 0.7 to 1.

\section{ $H(z)-z$ DATA AS  CONSTRAINING TOOL}

For a flat universe containing only radiation, cold matter (or baryon) and the MCG, the Friedmann
equation can be expressed as
\begin{equation}
H^{2}(H_{0},A_{s},B,\alpha,z)=H^{2}_{0}f^{2}(A_{s},B,\alpha,z)
\end{equation}
where, 
\begin{eqnarray}
f(A_{s},B,\alpha,z)=[\Omega_{r0}(1+z)^4+\Omega_{j0}(1+z)^3
+ \nonumber \\ (1-\Omega_{r0}- \Omega_{j0}) 
(A_s+(1-A_s)(1+z)^{3n})^{\frac{1}{1+\alpha}} ]^{\frac{1}{2}}
\end{eqnarray}
with $j=m$ for CDM model and $j=b$ for UDME model.
The best fit values for model parameters $A_{s}$, $B$, $\alpha$ and $H_{0}$
can be determined by minimizing as follows
\begin{equation}
\chi^{2}(H_{0},A_{s},B,\alpha,z)=\sum\frac{(H(H_{0},A_{s},B,\alpha,z)-H_{obs}(z))^2}{\sigma^{2}_{z}}.
\end{equation}

Since we are interested in determining the model parameters, $H_{0}$ is  not an important parameter here. So we marginalize over $H_{0}$ to evaluate the probability distribution 
function for  $A_{s}$, $B$, $\alpha$ as
\begin{equation}
L(A_{s},B,\alpha)=\int  \left[dH_{0}P(H_{0})\exp(\frac 
{-\chi^{2}(H_{0},A_{s},B,\alpha,z)}{2}) \right ]
\end{equation}

where $P(H_{0})$ is the prior distribution function for the present Hubble constant.
We consider Gaussian priors $H_{0}=72\pm 8 $. For $-1\sigma$ level calculation the limit of integration will be from 64 to 72 and for $1\sigma$ level calculation
the limit of integration becomes 72 to 80.
Minimizing $\chi^{2}$ determines the maximum $L(A_{s},B,\alpha)$ value. We determine  the maximum value of the function $L(A_{s},B,\alpha)$
by plotting the function with any of its parameter keeping other two fixed. As a result we get the maximum values of the parameters $A_{s}$, $B$ and $\alpha$.
Consequently a relation between $B$ and $A_{s}$ for various $\alpha$ can be established.

\input{epsf}
\begin{figure}
\begin{center}
\epsffile{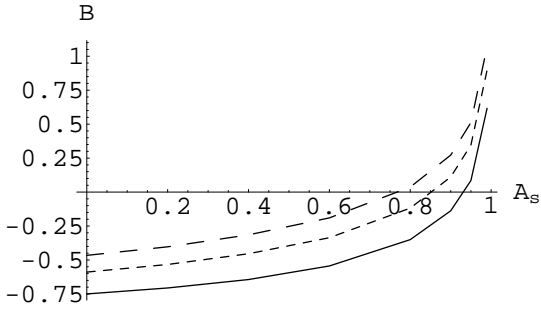}
\caption {CDM,Dashed line for 3$\sigma$ level,Dotted line for 2$\sigma$ level,Thin line for 1$\sigma$ level, 
for $\alpha = 0.01$}
\end{center}
\end{figure}

In CDM model, variation of $B$ with $A_s$ for $\alpha=0.01,\; 0.5,\;0.99$ at -1$\sigma$, -2$\sigma$ and -3$\sigma$ level respectively are shown in figs. (3)-(5). We note that 
as the value of $A_{S}$ tends to 1  we see that the $B$ parameter picks up
positive values (i) up to 1.07 (fig.3), (ii) upto 0.62 (fig.4), (iii) upto 0.36 (fig.5) in accordance with the $H(z)-z$ data. Thus as $\alpha$ increases, $B$ decreases. 
In UDME model variation of $B$ with $A_s$ for $\alpha=0,\; 0.5,\;1$  at $\pm$1$\sigma$, $\pm$2$\sigma$ and $\pm$3$\sigma$ level respectively are shown in figs. (6)-(8). We note that 
as the value of $A_{S}$ tends to 1  we see that the $B$ parameter picks up
positive values (i) upto 1.35 (fig.6), (ii) upto 0.84 (fig.7), (iii) upto 0.58 (fig.8) in accordance with the $H(z)-z$ data. Thus as $\alpha$ increases, $B$ decreases but compared to CDM model $B$ parameter values in UDME is more for a given $\alpha$ and $A_s$.

\input{epsf}
\begin{figure}
\begin{center}
\epsffile{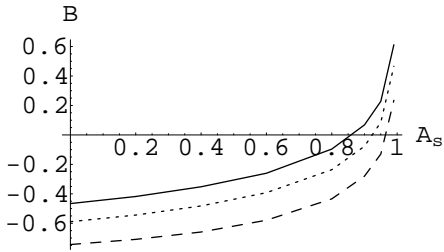}
\caption {CDM,Thin line for 3$\sigma$ level,Dotted line for 2$\sigma$ level,Dashed  line for 1$\sigma$ level, 
for $\alpha=0.5$}
\end{center}
\end{figure}

\input{epsf}
\begin{figure}
\begin{center}
\epsffile{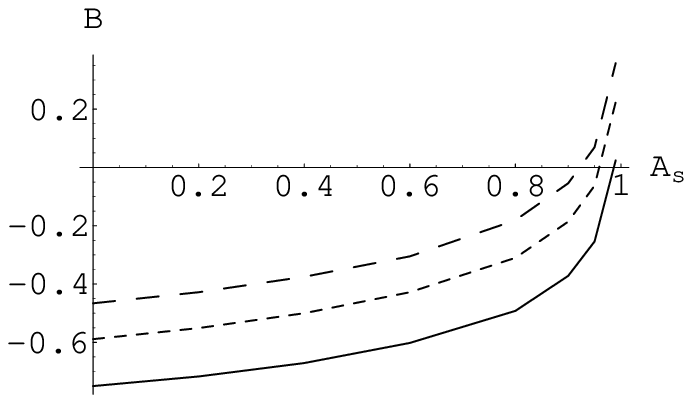}
\caption {CDM,Thin line for $3\sigma$ level,Dashed line for $2\sigma$ level,Dotted line for $3\sigma$ level for $\alpha=0.99$ }
\end{center}
\end{figure}

\input{epsf}
\begin{figure}
\begin{center}
\epsffile{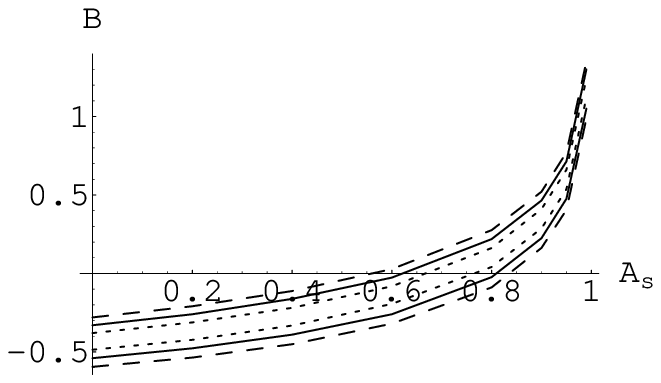}
\caption {UDME,Dotted pair of lines  represent boundary of 1$\sigma$,Thin line pair
represent 2$\sigma$,Dashed pair for 3$\sigma$ level,for $\alpha=0$}
\end{center}
\end{figure}

\input{epsf}
\begin{figure}
\begin{center}
\epsffile{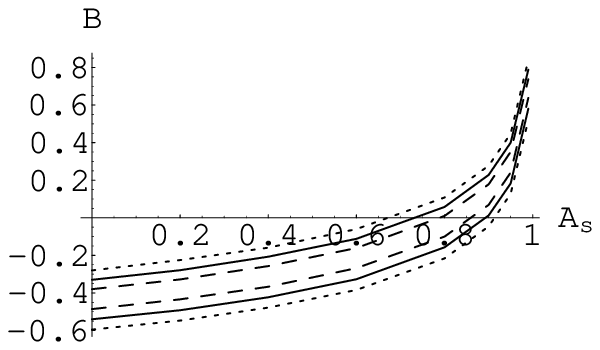}
\caption {UDME,Dashed pair of represent boundary of $1\sigma$,Thin line pair represent
$2\sigma$,Dotted pair for $3\sigma$ level,for $\alpha=0.5$ }
\end{center}
\end{figure}

\input{epsf}
\begin{figure}
\begin{center}
\epsffile{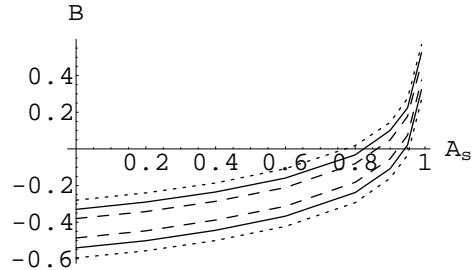}
\caption {UDME,Dashed pair of represent boundary of $1\sigma$,Thin line pair represent $2\sigma$,Dotted pair for $3\sigma$ level,for $\alpha=1.0$ }
\end{center}
\end{figure}

Thus in CDM the range  for $B$ is 0 to 1.07 and in Unified Dark Energy model the range is 0 to 1.35 upto 99.7 percent confidence level.

In CDM $B$ is positive only when  $A_{s}$ is within 0.76 to 1 for $\alpha$  0 to 1 and
in UDME $B$ is positive (so, permissible) only when  $A_{s}$ is within 0.57 to 1 for $\alpha$ lying between 0 to 1. 

\section{Viability of the model from union compilation of supernova magnitudes and redshift data}

We have already found the best-fit values of the parameters of MCG in our models from H(z)-z data part. For CDM model the best-fit values of the parameters are $A_{s}=.99$, $B=.01$, $\alpha=.01$ and  for UDME model it is $A_{s}=.8$,$B=.06$, $\alpha=.11$. In order to check its validity we used these best-fit values to find supernovae magnitudes at different redshift for the two models of our consideration. From there we have drawn supernovae magnitudes vs redshift curve using those best-fit values of the models. We compared these supernovae magnitudes vs. redshift curve of those two models to the original curve of union compilation data (Kowalaski {\it et al.}) (between those two parameters).
 Fig.9 shows a plot of $\mu (z)$ vs. $z$ obtained from the CDM model (the continuous line) along with that obtained from union compilation data (the dots). The same curve is drawn for the UDME model in Fig. 10 (continuous line for UDME model and dots for union compilation data). As one can see from these two curves, both the CDM and UDME models are in excellent agreement with union compilation data.

\input{epsf}
\begin{figure}
\begin{center}
\epsffile{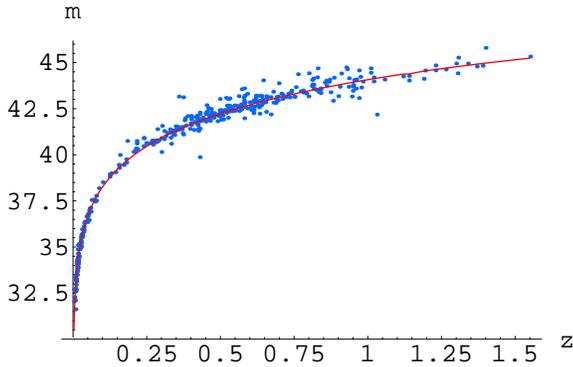}
\caption {$\mu (z)$ vs. $Z$ curves for CDM model and union compilation data (in the figure $m$ stands for $\mu (z) $) }
\end{center}
\end{figure}

\input{epsf}
\begin{figure}
\begin{center}
\epsffile{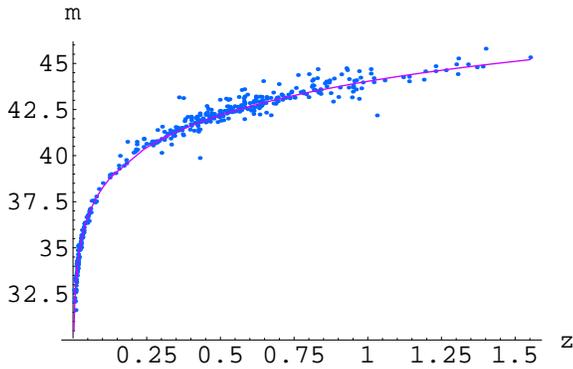}
\caption {$\mu (z)$ vs. $Z$ curves for UDME model and union compilation data (in the figure $m$ stands for $\mu (z) $). }
\end{center}
\end{figure}

\section{Discussion}

Cosmological models with Modified Chaplygin Gas  presented here, contain three different parameters, $\; A_{s}$, $ B$, $ \alpha $. From the age constancy we determine permissible range of values of $B$ parameter. In sec. 3 we plot $ B \; vs. \; A_s$ for different values of $\alpha$ in figs. (1) and (2).  The figures are plotted for both positive and negative values of $B$.  In the case of CDM we note that $B$ can pick up  positive values upto 0.2  for $.97 \leq A_{s} < 1 $,  $ 0 \leq \alpha \leq 0.39 $. However, for UDME model we note that $B$ can pick up positive values upto 1.02 for  $0.7 \leq A_{s} <  1 $, $0 \leq  \alpha \leq 1 $.

In sec. (4), using the  Hubble parameter vs. redshift data we obtain the constraints on $B$ using chi-square minimization technique. To obtain a viable cosmology with MCG we restrict to positive values of  $B$. The constraints on $B$ are   : (i) $ 0 \leq B \leq 1.07 $ for $0.76 \leq A_{s} < 1 $, and  $ 0 \leq \alpha \leq 1 $ in CDM and (ii) $ 0 \leq B \leq 1.35 $ for  $0.56 \leq A_{s} < 1 $,  $ 0 \leq \alpha \leq 1 $ for UDME. 
For UDME model the  range of values of $B$ is found to be more than that of CDM. If the age constant parameter is decreased then we note that the values of $B$  permitted by CDM and UDME  model are  in agreement with that found by chi-square minimization of the observed H(z) vs z data. Consequently the limiting value of the age of our universe is pushed to lower values ($t < 13.6 \; Billion \; years $).

The  best-fit values of the parameters obtained here for CDM and UDME models are in agreement with  union compilation data.  We note that the best-fit values of our models are $ A_{s}=0.99 $,$ B=0.01 $,$ \alpha=0.01 $ for CDM and  $A_{s}=0.8$,$ B=0.06 $,$ \alpha=0.11 $ for UDME model.

\section*{Acknowledgments}
PT and SG would like to thank {\it IUCAA Reference Centre}, Physics Department, N.B.U for extending the facilities of research work. SG would like to thank University of North Bengal for awarding Junior Research Fellowship.

\label{lastpage}

\end{document}